\documentclass[graphicx,reprint]{revtex4-1}
\usepackage{graphicx}
\usepackage{dcolumn}
\usepackage{sistyle}
\usepackage{upgreek}
\usepackage{amsmath}
\usepackage{xcolor}

\newcommand{\cnt}{\mathrm{\scriptscriptstyle CNT}}

\renewcommand{\d}{\mathrm{d}}

\begin{document}

\title{Adhesion energy of single wall carbon nanotube loops on various substrates}



\author{Tianjun LI}
\affiliation{Department of Physics, Shaoxing University - 508, Huancheng West Rd. Shaoxing, 312000, China}
\affiliation{Universit\'e de Lyon, Laboratoire de physique, ENS de Lyon, CNRS - 46, all\'ee d'Italie, Lyon 69364, France}
\author{Anthony AYARI}
\affiliation{Institut Lumi\`ere Mati\`ere, UMR5306 Universit\'e Lyon 1-CNRS, Universit\'e de Lyon 69622 Villeurbanne cedex, France.}
\author{Ludovic BELLON}
\email{ludovic.bellon@ens-lyon.fr}
\affiliation{Universit\'e de Lyon, Laboratoire de physique, ENS de Lyon, CNRS - 46, all\'ee d'Italie, Lyon 69364, France}

\date{\today}

\begin{abstract}
The physics of adhesion of one-dimensional nano structures such as nanotubes, nano wires, and biopolymers on different material substrates is of great interest for the study of biological adhesion and the development of nano electronics and nano mechanics. In this paper, we present force spectroscopy experiments of a single wall carbon nanotube loop using our home-made interferometric atomic force microscope. Characteristic force plateaux during the peeling process allows us to access to quantitative values of the adhesion energy per unit length on various substrates: graphite, mica, platinum, gold and silicon. By combining a time-frequency analysis of the deflexion of the cantilever, we access to the dynamic stiffness of the contact, providing more information on the nanotube configurations and its intrinsic mechanical properties.
\end{abstract}

\pacs{}

\maketitle 

\section{Introduction}
Since their discovery~\cite{Iijima1993}, carbon nanotubes (CNTs) have attracted the interests of scientist for their unique electrical~\cite{Yao2000}, thermal~\cite{Iijima1991} and mechanical properties~\cite{Lanzani2011} and are foreseen as a major material in a huge range of applications in the next few decades, from material reinforcement~\cite{Bal2007} to components of nanoscale electronics~\cite{Collins2000} and mechanics, for instance, nanoswitches~\cite{Cumings2000}, motors~\cite{Fennimore2003}, actuators~\cite{Ke2004,Breuer2004,Kis2008}, etc. They are also widely used as a bench system to study fundamental physical phenomena on the mesoscopic scale, and represent as such an archetype of nano-objects. Whatever exceptional their intrinsic properties are, their use in any application are linked to their interactions with their environment, mainly through the weak adhesive Van der Waals force. The study of the Van der Waals interaction between nanotubes and the rest of the world can thus help to understand the physics of polymer nanocomposites~\cite{Breuer2004,Hussain-2006,Bal2007}, setae adhesion of geckos~\cite{Correa-Duarte2004}, protein filament adhesion in mussels~\cite{Sever2004}, nanotube-tipped atomic force microscopy (AFM) probes~\cite{Strus2005}, nanoscale sensors~\cite{Mahar2007}, gecko-foot-mimetic dry adhesives~\cite{Qu2008}, etc.

Up to date, the adhesive properties of nanotube has been mostly probed by various smart but indirect measurement. Hertel and coworkers~\cite{Hertel1998,Hertel1998a} imaged by AFM the shape of crossed nanotubes adsorbed on a silicon substrate. The profile of the top nanotube balances the deformation energy and the surface energy lost in this configuration, giving a estimated adhesion energy when mechanical properties of nanotube are assumed. Kis and coworkers~\cite{Kis2006} performed a direct measurement using an AFM tip to pull the inner core of a telescopic multi wall CNT. Their experiment demonstrated a friction free interaction between the concentric layers, and provides an estimation of the adhesion for this very specific geometry and material.

However, in these experiments, one usually access either intrinsic properties of the nanotube, or its interaction with its environment using hypotheses on the other properties. Direct measurement of adhesive interaction can provide quantitative values of several properties in one single test and are thus an appealing method. A peeling test is a potentially powerful technique to characterize the adhesion properties of carbon nanotubes or nano wires with various substrates. A few experiments have been conducted~\cite{Ishikawa2008,Ishikawa2009,Ke2010,Strus2009,Xie2010}, along with theoretical/numerical modeling~\cite{Oyharcabal2005,Sasaki2008}. Quantitative measurement however have not been easily achieved, and experimental data analysis relies on complex comparison with numerical simulations. We recently proposed a simpler protocol to perform peeling test~\cite{Buchoux-2011}, which allow direct and quantitative characterization of the adhesion properties of nanotubes. A SWCNT is anchored to the tip of an AFM cantilever and is simply pushed almost perpendicularly to a flat surface. The VdW interaction causes part of the nanotube to be adsorbed to the surface when the induced bending is strong enough, and the analysis of the force curve leads both to quantitative information on the adhesion process and on the nanotube itself.

In this paper, we present an extension of this protocol to CNT loops on various substrates, leading to quantitative values of the energy of interaction of the nanotube with substrates of graphite, mica, platinum, gold and silicon. We first present the samples (CNT and substrates), and the acquired data during approach-retract cycles: force, dynamic stiffness, compression. We then give the analyzing framework, and discuss the nanotube configurations during peeling. We finally conclude, giving quantitative values of the adhesion energy of a SWCNT on the five different substrates.

\section{Experiments}

\begin{figure}
\centering
\includegraphics{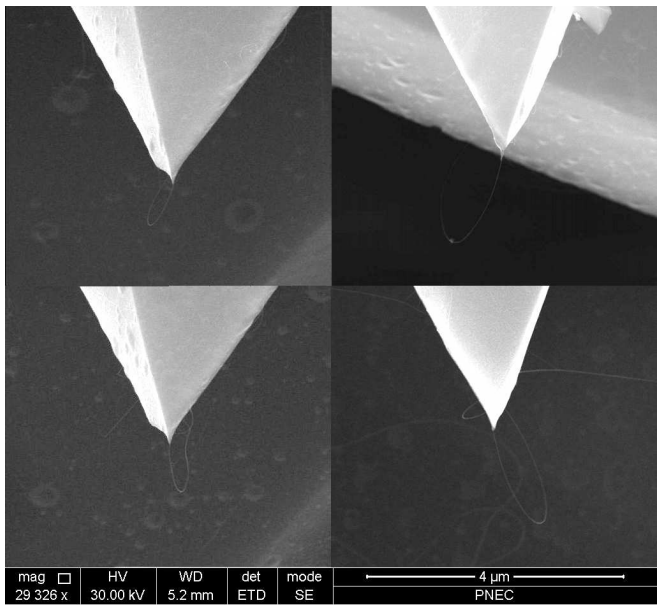}

\vspace{0.5cm}

\includegraphics{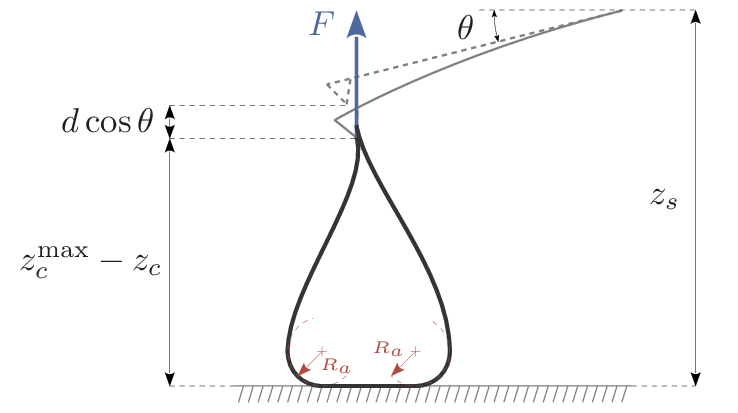}
\caption{(Color online) -- (Top) Scanning electron micrograph of several SWCNT grown directly on a AFM tip. Long nanotubes tend to form loops on the tip, with a typical diameter around $\SI{1}{\micro m}$. (Bottom) When the nanotube loop is pushed against a flat surface, part of the nanotube is adsorbed on the surface due to Van der Waals interactions. The radius of curvature $R_{a}$ at the last point of contact on the CNT with the substrate is fixed by an equilibrium between the adhesion of the adsorbed part and the bending of the free standing part of the nanotube. From the measurement of the AFM cantilever deflection $d$ and sample position $z_{s}$, the force $F$ acting on the nanotube and its compression $z_{c}$ can be recorded.}
\label{fig:SEM-sketch}
\end{figure}

\subsection{Sample preparation}

The nanotubes are grown directly~\cite{Marty2006} at the tip apex of AFM probes by Chemical Vapor Deposition (CVD): the bare silicon cantilevers are fully dipped into the catalyst solution, then gently dried in a nitrogen flux before being placed in the furnace. CNTs grow everywhere on the cantilever, and around 1 every 3 cantilevers present a CNT at the tip. The parameters are tuned to grow long SWCNT (a few micrometer), leading to a high probability of having nanotube loops (two anchoring points on the tip).  A few SEM images of our samples are presented on figure~\ref{fig:SEM-sketch}, note however that nanotubes are often detected during AFM measurements while they were not visible on the SEM images (the opposite is also true).

For the substrates, graphite and mica are first chosen as ideal candidates since a fresh layer is always easy to be cleaved before the test. The three other substrates we have chosen to investigate relative adhesion energies are surfaces of platinum (Pt), gold (Au) and silicon (Si). For the peeling tests, a very flat surface is indispensable so as to get rid of the impact from the morphology of the surface. We choose to use chips of commercial AFM cantilevers with coating of Pt, Au, and without coating (thus Si surface) as our substrates. Before the experiments, these 3 substrates undergo a treatment process in an ultrasound cleaner in ethanol for \SI{10}{mins}, isoproponal for \SI{10}{mins}, then in a plasma cleaner (medium power) for \SI{10}{mins} to avoid any contamination from the environment. They are finally kept in a clean and dry container for \SI{6}{hours} to eliminate any surface impact that might be changed by the cleaning process.

\subsection{Force curves}

\begin{figure}
\centering
\includegraphics{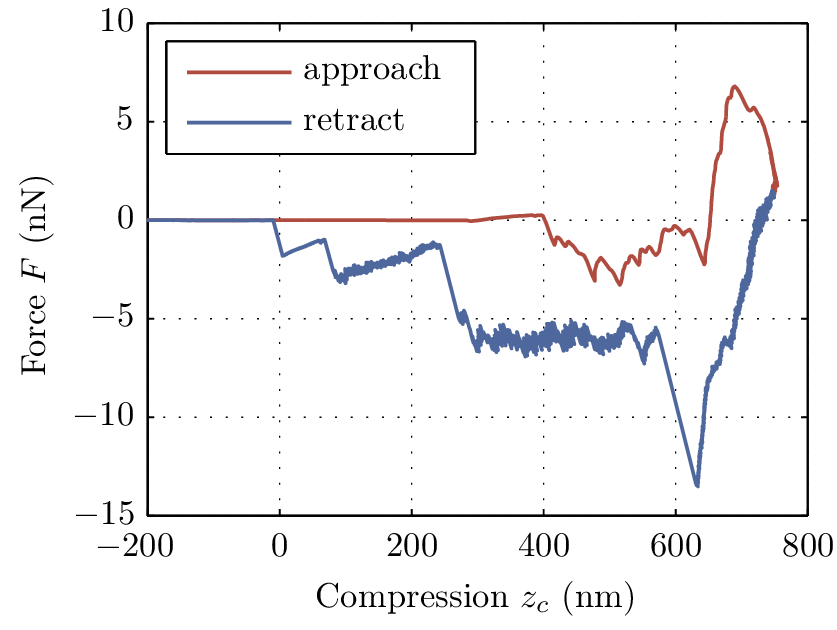}
\caption{(Color online) -- Force $F$ of a nanotube as a function of its compression $z_c$ on a graphite substrate. A strong hysteresis, due to the adhesion, can be noticed between approach (red, top curve) and retraction (blue, bottom curve). Force plateaux, characteristic of a peeling mechanism, are observed during retraction.}
\label{fig:forcecurve_graphite}
\end{figure}

In experiment, the nanotube loop is pushed against a flat sample, as shown in the schematic diagram of figure~\ref{fig:SEM-sketch}. The translation of the substrate in the experiment is performed with a piezo translation platform operated in closed loop, featuring an accuracy of \SI{0.3}{nm} rms (PI P527.3). We measure the deflexion $d$ of the AFM cantilever with a home-made highly sensitive quadrature phase interferometer~\cite{Schonenberger-1989,Bellon-2002-OptCom,Paolino-2013}. The deflexion $d$ and the sample vertical position $z_s$ are simultaneously recorded with high resolution acquisition cards (NI-PXI-4462) at $\SI{200}{kHz}$.

With both $z_s$ and $d$ being calibrated, using a proper definition of origins, we can compute at any time the compression of the nanotube:
\begin{equation}
z_{c}=z_{s}-d\cos\theta
\end{equation}
where $\theta=\ang{15}$ accounts for the inclination of the cantilever with the substrate. We can also compute the vertical force acting on the nanotube:
\begin{equation}
F=-\frac{k_{s}}{\cos\theta} d
\end{equation}
with $k_{s}$ the static stiffness of the AFM cantilever (calibrated from its thermal noise~\cite{Butt-1995}). Using compression instead of sample position allows us to take into accounts the compliance of the cantilever, thus to focus on the nanotube properties only in the force versus compression curves. An example of such force curve is plotted in figure~\ref{fig:forcecurve_graphite} for a substrate of graphite. A strong hysteresis can be noted during the approach-retract cycle, pointing at the nanotube changing its configuration during compression. The force is mostly attractive (except at the end of the approach), hinting at adhesion as the main interaction process between the nanotube and the substrate. Finally, we note long force plateaux during the retraction: this is the signature of a peeling process~\cite{Buchoux-2011}.

Let us summarize the analysis presented in ref.~\cite{Buchoux-2011} to give grounds for this last claim. We denote by $E_a$ the energy of adhesion per unit length of the nanotube on the surface. As soon as part of the nanotube is adsorbed, the systems tends to minimize its energy by maximizing the absorbed length. However, this process increases the bending of the free standing part of the nanotube and the associated curvature energy. The adsorbed length is thus a balance between adhesion and bending, which leads to a constant radius of curvature $R_a$ at the contact point. If the free standing part of the nanotube is long enough compared to $R_a$, its shape does not change much when it is being peeled from the surface. The vertical displacement $\delta z$ needed to peel a small length $\delta l$ from the surface is thus the same length: $\delta z \simeq \delta l$. In quasi static displacement, the work produced by the pulling force $F\delta z$ is thus equivalent to the energy released $- 2 E_a\delta l$, leading to $F \simeq - 2 E_a$ (the factor 2 here accounts for the loop geometry of the CNT : we peel the same length $\delta l$ for the two strands). Peeling a nanotube loop from a surface thus results in a flat force-compression curve, and the value of the force plateau gives directly the value of adhesive energy per unit length. For the longest plateau (A) of this is nanotube on graphite, we read for example $E_a \simeq \SI{3}{nJ/m}$.

For every CNT and substrate, we perform at least 50 approach-retract cycles: during the first contacts of the nanotube, its configuration can change significantly before it reaches a stationary behavior. Some nanotubes can be lost during the process, other never reach a stable operation state. From the grown CNTs, we select only those having clear peeling process signature: long force plateaux, force curve reproducible on various substrate, stable in time. Most indeed have many defects resulting in a too complex interpretation of the data. We present here the results corresponding to a single nanotube on various substrate, though we did observe similar behavior for other samples. No SEM image of this nanotube in its useful configuration could taken : we were unable to see it under the electron beam, though the force signature is the best we observed !

\subsection{Dynamic stiffness\label{dynamicstiff}}

\begin{figure}
\centering
\includegraphics{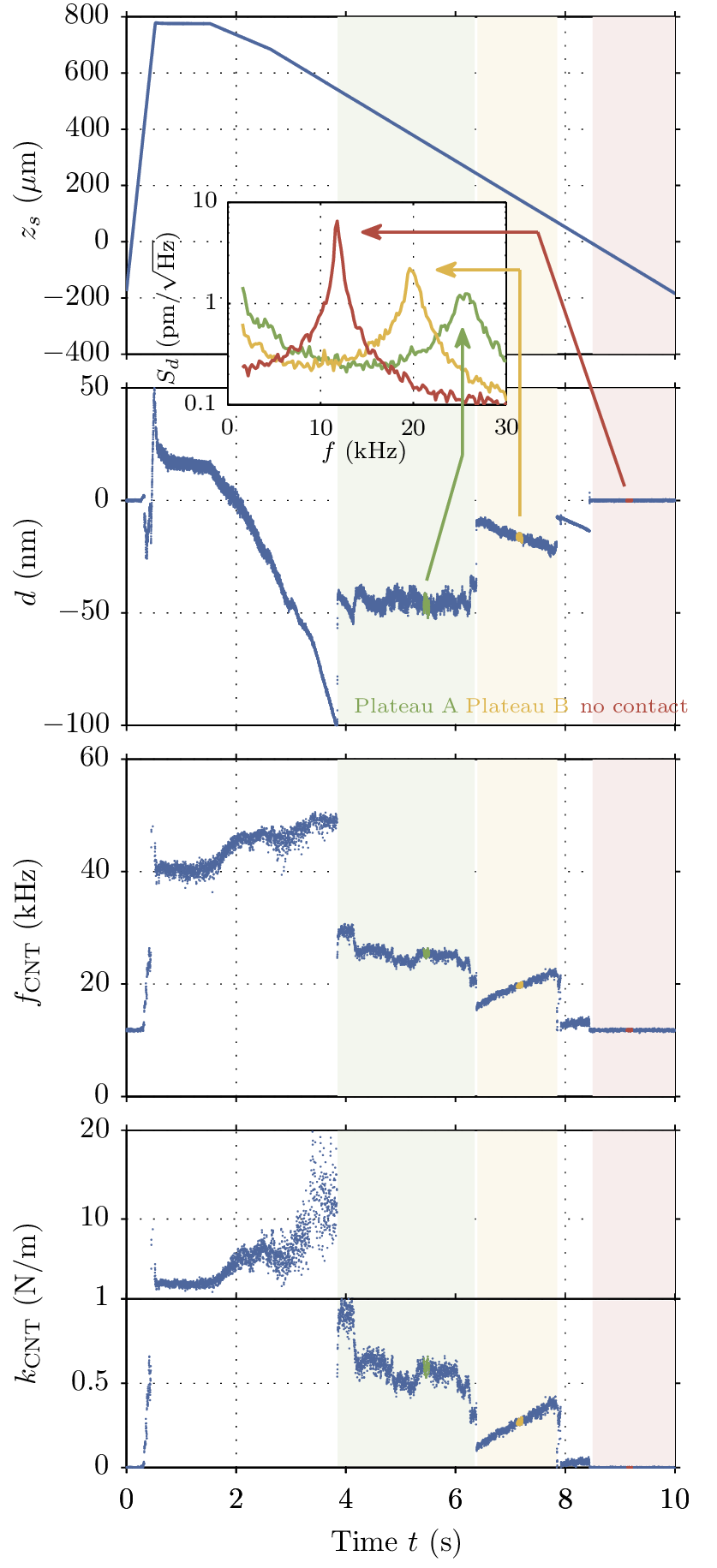}
\caption{(Color online) -- Time evolution of the sample position $z_s$, deflexion $d$, resonant frequency $f_{\cnt}$ and dynamic stiffness of the nanotube in contact $k_{\cnt}$ during a slow approach-retract cycle (note the dual vertical scale of this last plot). The power spectrum density (PSD) $S_{d}$ of the fluctuation of deflexion, computed in a $\SI{0.1}{s}$ window around $t=\SI{5}{s}$, $t=\SI{7}{s}$ (peeling configurations) and $t=\SI{9}{s}$ (no contact), is shown in the inset. The shift of the resonant frequency of the oscillator (cantilever + nanotube in contact) can be used to compute the dynamic stiffness of the contact (see figure~\ref{fig:freqshiftth}). A time-frequency analysis allows us to track this frequency shift as a function of time, following the maxima of the PSD computed in every $\SI{5}{ms}$ windows.}
\label{fig:timefreq}
\end{figure}

Following refs.~\cite{Buchoux-2009,Buchoux-2011}, we analyse during the same approach-retract cycles the dynamic properties of the nanotube-substrate contact by following the evolution of the thermal noise spectrum during the peeling process. Indeed, when the nanotube is in contact with the sample, the contact stiffness sums with the cantilever spring constant and leads to a higher resonance frequency of the system. The fluctuations of the deflection, driven by the random thermal noise excitation of the oscillator, allow us to track this frequency shift during contact, as illustrated in the inset of figure~\ref{fig:timefreq}: the power spectrum density (PSD) of the deflection $S_{d}$ presents the characteristic lorenzian shape of the thermal noise of a simple harmonic oscillator, peaked at frequencies ranging from $\SI{12}{kHz}$ (free cantilever, no contact), to $\SI{26}{kHz}$ (peeling configuration around time $t=\SI{5}{s}$).

\begin{figure}
\centering
\includegraphics{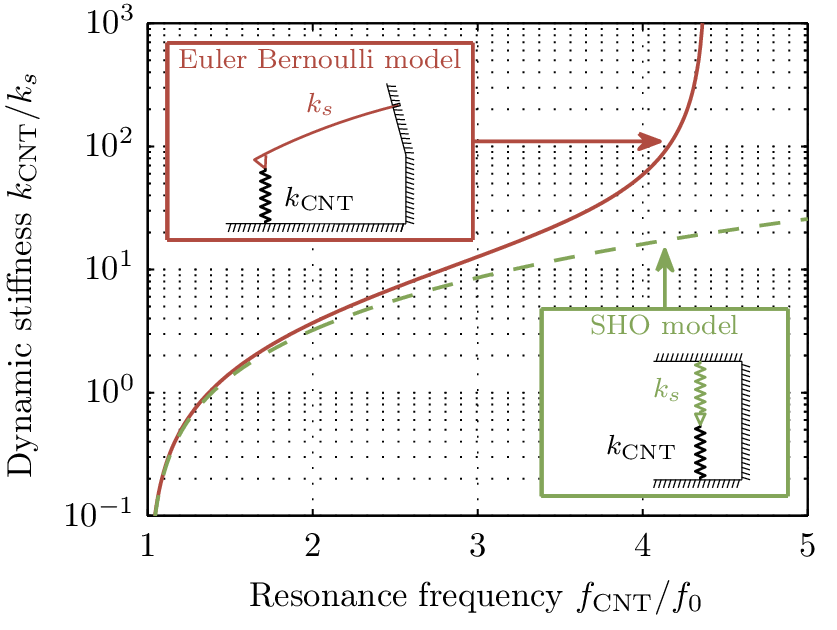}
\caption{(Color online) -- Dynamic stiffness $k_{CNT}$ deduced from the resonance frequency $f_{CNT}$ of the oscillator: when the dynamic stiffness is not too large compared to the static stiffness $k_{s}$ of the free cantilever, a simple harmonic oscillator (SHO) model with a spring is sufficient, but for larger values of $k_{\cnt}$, a full Euler-Bernoulli model of the cantilever is required: a hard contact produces a finite frequency shift ($f_{\cnt}/f_{0}\approx4.385$ when $k_{\cnt}\rightarrow\infty$).}
\label{fig:freqshiftth}
\end{figure}

From this frequency shift, we can recover the equivalent stiffness $k_{\cnt}$ of the CNT in contact with the substrate:
\begin{equation} \label{eq:fshift0}
k_{\cnt}=\frac{k_{0}}{\cos^2\theta} \left[\left(\frac{f_{\cnt}}{f_0}\right)^2-1\right]
\end{equation}
where $f_{0}$ and $f_{\cnt}$ are respectively the resonant frequency out of and in contact, and $k_{0}$ is the free cantilever dynamic stiffness. In this formula, we approximate the cantilever by a simple spring $k_{0}$, however when the frequency shift is significant as in this measurement, a complete Euler-Bernoulli description of the cantilever is better suited. It leads to the following relation between the dynamic contact stiffness and the resonance frequency of the first mode of oscillation~\cite{Buchoux-2009}:
\begin{equation} \label{eq:fshiftEB1}
k_{\cnt} = \frac{k_{s}}{\cos^{2} \theta} \frac{\alpha^{3} (1+\cos \alpha \cosh \alpha)}{3 (\cos \alpha \sinh \alpha - \sin \alpha \cosh \alpha)}
\end{equation}
with
\begin{equation} \label{eq:fshiftEB2}
\alpha=\alpha_{1} \sqrt{\frac{f_{\cnt}}{f_0}}
\end{equation}
where $\alpha_{1}\approx 1.875$ is the first spatial eigenvalue of a clamped-free Euler-Bernoulli mechanical beam.
In figure \ref{fig:freqshiftth}, we plot both relations \ref{eq:fshift0} and \ref{eq:fshiftEB1}: the simple spring approximation is valid till $f_{\cnt}\approx 2 f_{0}$, but the full relation is better suited as soon as the contact stiffness is larger than the cantilever one.

To have a continuous information about the contact stiffness at any compression, we perform a time-frequency analysis of the deflexion $d$, illustrated in figure~\ref{fig:timefreq}. The PSD of the deflexion $S_{d}$ is computed in every \SI{5}{ms} time window, corresponding to a \SI{0.6}{nm} sample displacement during retraction. We then extract the resonant frequency $f_{\cnt}$, corresponding to the maximum of the PSD. From equations~\ref{eq:fshiftEB1} and \ref{eq:fshiftEB2}, we finally access the dynamic stiffness $k_{\cnt}$ of the nanotube in contact at any time. At the maximum compression of the nanotube, we see from this analysis that we reach a hard contact between the AFM tip and the substrate: $k_{\cnt}$ reaches huge values (several $\SI{}{N/m}$) compared to the cantilever stiffness ($k_{s}=(0.121Â±0.005)\,\SI{}{N/m}$) when $f_{\cnt}/f_{0}\approx 4.4$.

\subsection{Mean interaction curves for the five substrates}

\begin{figure}
\centering
\includegraphics[width=0.45\textwidth]{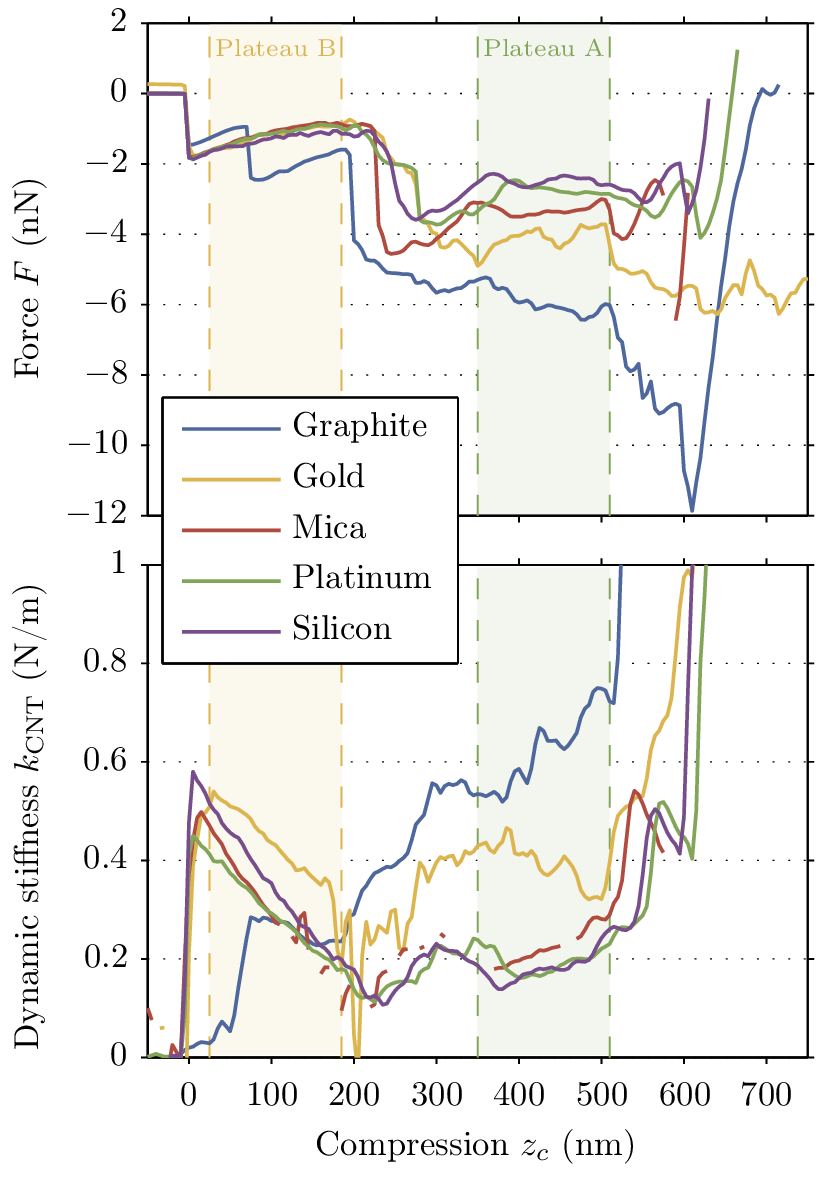}
\caption{(Color online) -- Mean force $F$ and dynamic stiffness $k_{CNT}$ as a function of compression $z_{c}$ measured during retraction on five different substrates: graphite, gold, mica, platinum and silicon. Each curve is the average of a minimum of 10 approach-retract cycles. Two ranges of compression corresponding to force plateaux are defined : plateau A for $z_{s}=[350-510]\,\SI{}{nm}$ and plateau B for $z_{s}=[35-185]\,\SI{}{nm}$. The zero of compression $z_{s}$ is defined by the loss of contact between the nanotube and the substrate.}
\label{fig:FandkCNTvscompression}
\end{figure}

For every substrate, we perform at least 10 very slow approach-retract cycles, and compute for each the force $F(z_{s})$ and dynamic stiffness $k_{\cnt}(z_{s})$ versus compression curves during the cycles. We then average these curves for each sample, and plot them in figure~\ref{fig:FandkCNTvscompression}. We focus here on the retractions only, which allow to probe the properties of the nanotube in a peeling configuration. The same nanotube has been used for all the substrates.

The curves for the five different samples have the same generic features : a steep force versus compression dependence and a huge dynamic stiffness at large compression, corresponding to a hard contact between the tip and the surface, then two distinct force plateaux (labelled A and B). The overall behavior is quite different in these two  ranges of compression. For the first plateau (A), $k_{\cnt}$ is rather flat on plateau A, with higher values corresponding to the samples with the higher adhesion force. On plateau B, the force is also rather flat, but it almost does not depend on the sample (with the exception of graphite). The stiffness presents a significant dependence on compression, with the highest values just before loosing contact. Again $k_{\cnt}$ does not present a important dependence on the substrate on plateau B. The shift between the 2 plateaux occurs at slightly different compressions for the five samples, the zero of compression $z_{c}$ being defined by the last contact point during retraction

\section{Adhesion energy and mechanical properties of the nanotube}

To understand the observed behavior, we use the framework of ref.~\cite{Buchoux-2011}, but we now consider a nanotube loop. In such a case, the CNT is indeed tangent to the surface before any contact and absorption is even more relevant than for a straight nanotube almost perpendicular to the surface. The first contact immediately leads to an absorption of the nanotube. The loss of contact also starts directly from the absorbed state, with no intermediate situation where only the tip of a straight nanotube is in contact. The behavior of the absorbed loop is however equivalent to having two straight nanotubes absorbed in parallel, and results of ref.~\cite{Buchoux-2011} are directly applicable, except that the force and contact stiffness should be divided by 2 for each strand of the nanotube !

Each part of the CNT in contact with the sample is thus described as an elastic line, incompressible along its axis. The shape of the line is given by a balance between adhesion and curvature, leading to the radius of curvature $R_{a}$ at the last contact point before the free standing part :
\begin{equation}
R_{a}=\sqrt{\frac{EI}{2E_{a}}}
\end{equation}
where $E_{a}$ is the energy of adhesion per unit length, $E$ the nanotube Young's modulus, and $I=\pi d_{\cnt}^{3} t_{\cnt}/8$ its quadratic moment ($d_{\cnt}$ its diameter, $t_{\cnt}$ the wall thickness --- $\SI{0.34}{nm}$ for a SWCNT~\cite{Kis2008}). As long as the free standing part of the nanotube is much larger than $R_{a}$, if we neglect horizontal components of the interaction, the force of interaction with the substrate should be constant and equal to $- E_{a}$~\cite{Buchoux-2011}, hence
\begin{equation}
F= - 2 E_{a}
\end{equation}
for a CNT loop.

The static stiffness, defined as $\d F / \d z_{c}$ is thus zero on the force plateau. However, as clearly illustrated in figure~\ref{fig:timefreq} or \ref{fig:FandkCNTvscompression}, the dynamic stiffness measured at the resonant frequency of the oscillator (cantilever + CNT in contact) is not zero: at a few tens of kHz, adhesion has no time to switch on and off for the small thermal fluctuations, and the absorbed part of the CNT can be considered as rigidly clamped to the substrate. In such a case, using the same hypotheses (long nanotube, negligible horizontal forces, 2 strands), we compute for the dynamic stiffness~\cite{Buchoux-2011}:
\begin{equation}\label{kcntra}
k_{\cnt}=2(1+\sqrt{2})\frac{E_a}{R_a}=8(1+\sqrt{2})\sqrt{\frac{E_{a}^{3}}{\pi E d_{\cnt}^{3} t_{\cnt}}}
\end{equation}
The dynamic stiffness should therefore present a plateau in correspondance to the force one, with values scaling as $E_{a}^{3/2}$.

\begin{figure}
\centering
\includegraphics{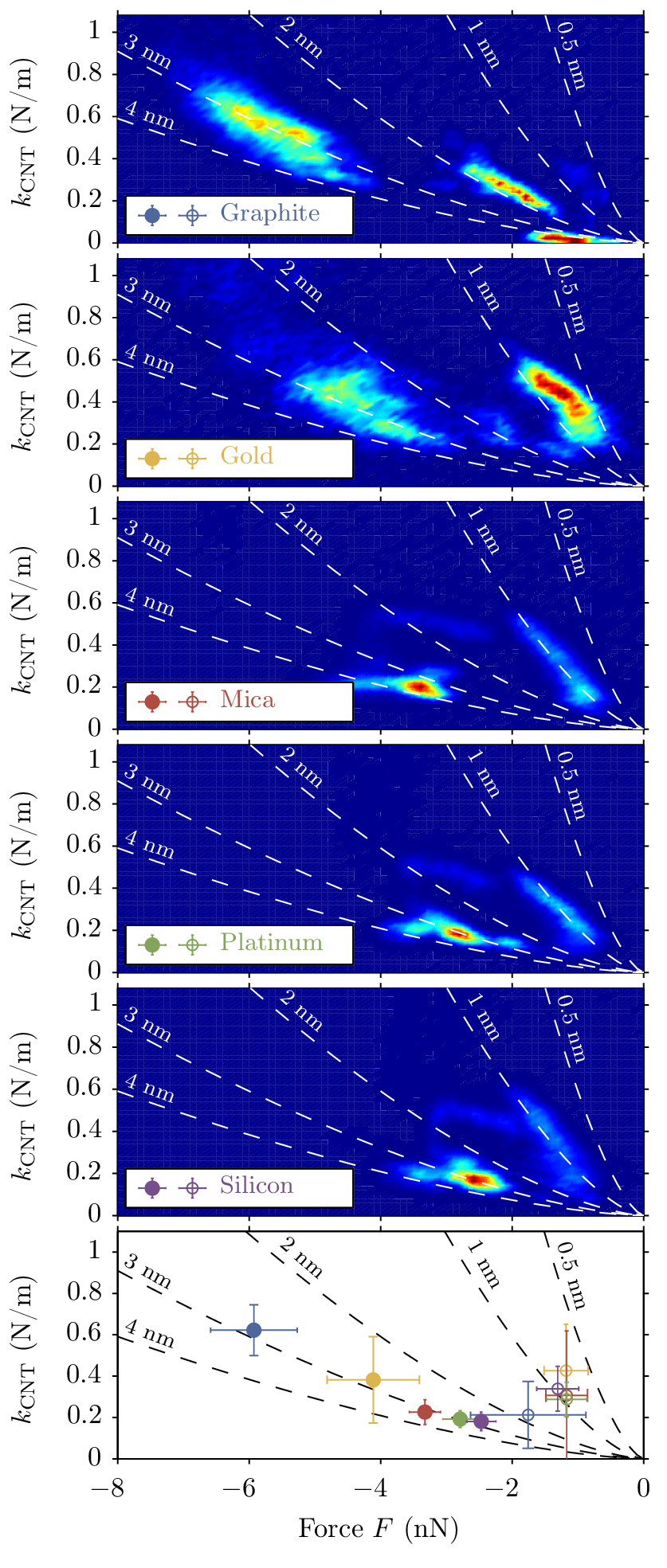}
\vspace{-0.5cm}
\caption{(Color online) -- To understand the configurations of the CNT during retraction, we plot the joint histograms of $k_{\cnt}$ and $F$, for the five substrates. The relation expected between $k_{\cnt}$ and $F$ for absorbed SWCNT loops of various diameters $d_{\cnt}=[0.5,1,2,3,4]\,\SI{}{nm}$ are superposed on the graphs. The nanotube explore mainly 2 different configurations : one at low force-low stiffness (plateau B), one at higher force and stiffness (plateau A). The mean value and standard deviation of $k_{\cnt}$ and $F$ on the two plateaux are plotted in the bottom graph (plain circles for A, empty circles for B). The configuration of plateau A agrees well with the model, with an estimated diameter $d_{\cnt}\approx\SI{3}{nm}$ for all the substrates. The configuration of plateau B would correspond to $d_{\cnt}\approx\SI{1}{nm}$, but show no substrate dependence and is attributed to adsorption on the AFM tip.}
\label{fig:kCNTvsF}
\end{figure}

To test those ideas, we compute for the five substrates the joint histograms of $k_{\cnt}$ and $F$ during all the retractions, and plot them in figure~\ref{fig:kCNTvsF}. As clearly illustrated in this figure, the CNT adopt two different configuration during peeling, corresponding to plateaux A and B. We also report in the bottom plot of figure~\ref{fig:kCNTvsF} the mean values of $k_{\cnt}$ and $F$ on the 2 plateaux defined in figure~\ref{fig:FandkCNTvscompression} for each sample. The data for plateau A are consistent with our model of peeling a CNT in the direction normal to the substrate : constant force $F$, constant dynamic stiffness $k_{\cnt}$, both depending on the substrate, but all sharing the same nanotube diameter : $d_{\cnt}\approx \SI{3}{nm}$.

The behavior on plateau B is not well described by the model:  except for graphite which present 2 different configurations during this range of compression, both $F$ and $k_{\cnt}$ depend on compression but not on the substrate. This can be the signature of an adhesion process of the nanotube loop on the silicon tip. Indeed, the two nanotube strands are not perfectly at the tip apex (or the hard contact we reach during the cycle would certainly break their clamping), so they can present some absorption-desorption mechanism. In such case, the signature should be quite independent of the substrate on the other side of the nanotube. Moreover, the peeling of the nanotube from the AFM tip is not normal to the surface of the silicon tip, so our model may not be suited: the force component parallel to the surface is not negligible any more, so the force and dynamic stiffness can depend on compression. The case of the graphite substrate, implying larger forces, apparently leads to different nanotube configurations in the range of compression corresponding to plateau B.

\begin{table}
\caption{(Color online) -- \label{table:Ea}Adhesion energies of a SWCNT on various substrates. The values of $E_{a}$ may differ for different nanotube parameters (diameter, chirality, purity, additional amorphous carbon layer), but not their relative values. The uncertainty corresponds to the standard deviation in the range of compression corresponding to plateau A for all forces curves on each substrate.}
\begin{tabular}{|l|l|}
\hline
Substrate & $E_{a}$ $(\SI{}{nJ/m})$ \\
\hline
Graphite & $2.96\pm0.33$ \\
\hline
Gold & $2.06\pm0.35$ \\
\hline
Mica & $1.66\pm0.12$ \\
\hline
Platinum & $1.39\pm0.13$ \\
\hline
Silicon & $1.24\pm0.11$ \\
\hline
\end{tabular}
\end{table}

We finally focus of the range of compression corresponding to plateau A only. The observations of the force and dynamic stiffness plateaux are well described by a peeling configuration, corresponding to a nanotube loop with $d_{\cnt}=\SI{3}{nm}$. The corresponding values of the adhesion energies of the five substrates are reported in table~\ref{table:Ea}: graphite has the stronger interaction with the CNT at $(2.96\pm0.33)\,\SI{}{nJ/m}$, while silicon has the lowest at $(1.24\pm0.11)\,\SI{}{nJ/m}$. Those numbers are in the same order of magnitude has other reported in the literature~\cite{Girifalco-2000,Hertel1998,Hertel1998a,Chen-2003,Kis2006,Strus2008,Strus2009CST,Ishikawa2008,Ishikawa2009,Ke2010}, though we notice a factor 3 with our previous measurements on graphite and mica~\cite{Buchoux-2011}. However, the nanotubes of those 2 sets of experiment were grown at different time in different laboratories, and may have some different adhesion properties due to different chiralities, amount of amorphous carbon around the SWCNT, etc. The ratio between the energy of adhesion are anyway preserved (nanotubes stick twice more on graphite than mica), so the relative values of substrates presented in table~\ref{table:Ea} are of broad utility.

\section{Conclusion}

In this article, we present some experiments where a SWCNT loop, grown directly on an AFM tip, is pushed against various substrates. The adhesion force and dynamic stiffness of the nanotube in contact with the sample are recorded as a function of its compression. The experimental data are analyzed in the framework of an elastic loop absorbed on a flat surface. During retraction, we observe force plateaux characteristic of a peeling mechanism. The cross information between force and dynamic stiffness helps to understand the configuration of the nanotube during the peeling process: some of the plateaux are attributed to adhesion on the substrate, while others are hinting at adsorption on the AFM tip. Quantitative values are derived for the nanotube diameter, and its energy of adhesion per unit length on various substrates: graphite, gold, mica, platinum and silicon.

Our experiments illustrate how CNT loops are useful nano-objects to probe peeling processes at the nanoscale, leading to quantitative measurement of their van der Walls interaction. This work provides an interesting insight into the physical mechanism of adhesion, and should be helpful in the design nanotube-based nanomechanical devices: if adhesion is (sought for clamping purposes for example), graphite is a good candidate, whereas is low interaction can be obtained using a silicon substrate. Other materials of technological relevance could be characterized using our simple protocol.

\begin{acknowledgments}
We thank F. Vittoz and F. Ropars for technical support, L. Champougny, M. Geitner, A. Petrosyan, J.P. Aim\'e and Z. Sun for stimulating discussions. This work has been supported by the ANR project \emph{HiResAFM} (ANR-11-JS04-012-01) of the Agence Nationale de la Recherche in France. Finally, T. J. LI would give thanks to Chinese Scholar Council (CSC) for financial support. The authors acknowledge the Plateforme Nanofils et Nanotubes Lyonnaise of the University Lyon1.
\end{acknowledgments}

\bibliography{CNTAdhesionVariousSubstrates}

\end{document}